\documentclass[10pt,aps,prb,twocolumn,superscriptaddress]{revtex4}

\usepackage{epsfig}
\usepackage{dcolumn}
\usepackage{bm}
\usepackage{amsmath}
\usepackage{amsfonts}
\usepackage{latexsym}
\usepackage{amssymb}
\usepackage{color}
\usepackage{hyperref}
\usepackage[normalem]{ulem}

\newcommand{\g}[1]{{\bf #1}}

\newcommand{\ug}{UGe$_2$}

\newcommand{\lcg}{La$_5$Co$_2$Ge$_3$}

\newcommand{\be}{\begin{equation}}
\newcommand{\ee}{\end{equation}}
\newcommand{\bea}{\begin{eqnarray}}
\newcommand{\eea}{\end{eqnarray}}
\newcommand{\ba}{\begin{eqnarray*}}
\newcommand{\ea}{\end{eqnarray*}}

 \usepackage{bbm}
\usepackage{bbold}

\usepackage{tikz,xcolor,hyperref}
\definecolor{lime}{HTML}{A6CE39}
\DeclareRobustCommand{\orcidicon}{%
	\begin{tikzpicture}
	\draw[lime, fill=lime] (0,0)
	circle [radius=0.16]
	node[white] {{\fontfamily{qag}\selectfont \tiny ID}};
	\draw[white, fill=white] (-0.0625,0.095)
	circle [radius=0.007];
	\end{tikzpicture}
	\hspace{-2mm}
}

\foreach \x in {A, ..., Z}{%
	\expandafter\xdef\csname orcid\x\endcsname{\noexpand\href{https://orcid.org/\csname orcidauthor\x\endcsname}{\noexpand\orcidicon}}
}

\begin{document}

\title{Spatially modulated, orbital selective ferromagnetism in La$_5$Co$_2$Ge$_3$} 
\author{Giuseppe Cuono\orcidA}
 \email{gcuono@magtop.ifpan.edu.pl}
\author{Carmine Autieri\orcidB}
\author{Marcin M. Wysoki\'nski\orcidC}
 \email{wysokinski@magtop.ifpan.edu.pl}
\affiliation{International Research Centre MagTop, Institute of
   Physics, Polish Academy of Sciences,\\ Aleja Lotnik\'ow 32/46,
   PL-02668 Warsaw, Poland}

\date{\today}

\begin{abstract}
 We present density functional theory calculations for low-$T_c$ metallic ferromagnet  \lcg\ at ambient and applied pressures. Our investigations reveal that the system is a quasi-one-dimensional ferromagnet with a peculiar coexistence of two different orbital-selective magnetic moments at two crystallographically inequivalent cobalt atoms, Co1 and Co2. Namely, due to different crystal-field splitting, the magnetic moment of Co1 atoms predominantly derives from $d_{xz}$ orbital whereas of Co2 atoms from $d_{xy}$ orbital. Consequently, Co1 and Co2 atoms develop unequal net magnetic moments, a feature that  gives rise to a periodic, spatial modulation of magnetization  along crystallographic $c$-direction.  
 The amplitude of the spatial modulation, small at ambient pressure, drastically increases with applied pressure, until Co2 atoms become nonmagnetic.
With the help of a toy model mimicking found orbital-selective ferromagnetic order, we demonstrate that the increasing amplitude of spatial modulation provides a consistent interpretation to the recently observed  resistivity anomaly  emerging at applied pressure identified as the appearance of  the {\it new state}. 
Although,  proposed here structural origin of the  spatial modulation of magnetic moments in \lcg\ is an alternative one to the advocated for this material ferromagnetic quantum criticality avoidance, the   effects of   quantum fluctuations can still play an important role at pressure larger than up-to-date measured 5GPa.     
\end{abstract}

\maketitle
\section{Introduction}

Low-$T_c$ metallic ferromagnets are systems in which quantum fluctuations can play a key role in understanding of their macroscopic properties when tuned with clean, i.e. not introducing disorder, means such as pressure or magnetic field \cite{Brando2016}. Namely, it has been theoretically predicted that ferromagnetic quantum critical point (QCP) in these systems is generically avoided and at low temperature a first order transition \cite{Belitz1997} or other, likely spatially modulated  phases \cite{Chubukov2004,Conduit2009,Karahasanovic2012}  should appear  instead. Moreover, quantum-fluctuation driven pressure-magnetic-field-temperature phase diagrams of low-$T_c$ itinerant ferromagnets can have a so-called tricritical wing structure \cite{Belitz2005}.   

Multiple experiments on metallic ferromagnets have shown consistency between measured properties and theoretical expectations relying on avoided QCP scenario. Namely, at low temperatures there have been observed: ({\it i}) change from a second to first-order ferromagnet (FM) to paramagnet  transition \cite{Pfleiderer2002,Uhlarz2004,Araki2015,Shimizu2015}, ({\it ii}) emergence of spatially modulated phase \cite{CeAgSb2,CeRuPO2,Taufour2016,Taufour2017,CeTiGe3,Gati2021} and ({\it iii}) appearance of tricritical wings \cite{Uhlarz2004,Taufour2010,Kotegawa2011,Aoki2011,Araki2015,Shimizu2015,Taufour2017}. Very recently itinerant ferromagnet \lcg, that is the focus of the current work, has joined the above list as an example of emergence of the new, potentially spatially modulated, phase out of FM under pressure \cite{Canfield2021}. 

For the proper description of metallic ferromagnets there has been also proposed a Stoner type of approach based on common among them electronic structure property 
that is mixing of correlated and uncorrelated orbitals in the vicinity of the Fermi level.
This approach, just as QCP avoidance approach, is able to rationalize observations of  first order transition \cite{Wysokinski2014R,Wysokinski2015R}, tricritical wings \cite{Wysokinski2015R,Abram2016}, and spatially modulated phase \cite{Wysokinski2019} under pressure. Although more restricted by the material-specific input parameters,  this microscopic approach has an advantage  as it captures the appearance of the two distinct FM phases ubiquitous among metallic ferromagnets \cite{Uhlarz2004,Pfleiderer2002,Kobayashi2007,Taufour2010,Taufour2017}. Moreover, the  approach  inspired the proposal of a consistent theory of the  superconductivity in \ug. \cite{Kadzielawa2018}

Possible ambiguity in the microscopic foundations of generic features among metallic ferromagnets calls for   critical 
 and thorough theoretical investigations of each of the compounds separately in order to provide a conclusive interpretation of particular observations. In this light recently reported measurements for low-$T_c$ metallic  ferromagnet \lcg\  indicating the appearance of a partially gapped, presumably due to an emergent antiferromagnetic (AFM) component, high-pressure magnetic {\it new state} \cite{Canfield2021},   raised a question about the underlying mechanism responsible for this novel phase.  
Therefore, in the current work, by means of  the density functional theory (DFT) calculations we analyze electronic and magnetic properties of \lcg\ in order to shed  a light on the origin of this enigmatic  high pressure magnetic phase. 

\begin{figure}[t]
	\centering
	\includegraphics[scale=0.405]{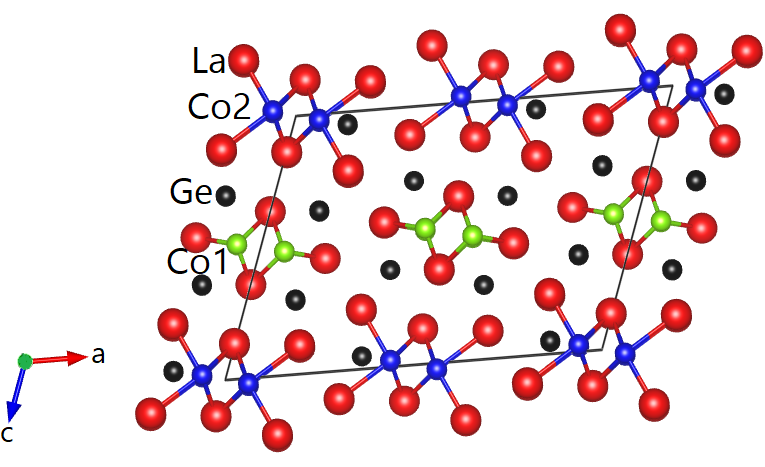}
	\includegraphics[scale=0.405]{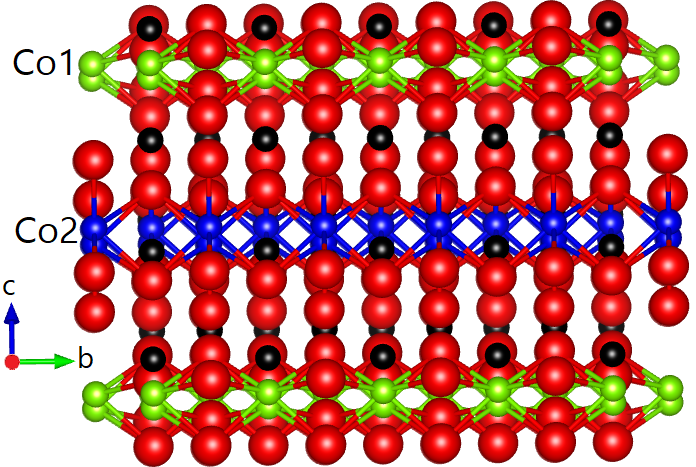}
	\caption{Crystal structure of La$_5$Co$_2$Ge$_3$. 
	Green and blue spheres denote Co1 and Co2 atoms, while red and black spheres La and Ge atoms.
	Co1 and Co2 atoms form chains along $b$-direction.
	}
	\label{Structure}
\end{figure}

Our results confirm that the high pressure phase in \lcg\ as suggested in Ref. \onlinecite{Canfield2021} is indeed characterized with a spatial modulation of magnetic moments. However,  the origin of a spatial modulation according to our calculations is structural and attributed to the $d$-orbital selectivity stemming from a presence of  crystallographically inequivalent cobalt atoms, Co1 and Co2 (see Fig. \ref{Structure}), rather than to purely electronic-like mechanism as in QCP avoidance \cite{Chubukov2004,Conduit2009,Karahasanovic2012} or correlated-uncorrelated orbitals mixing  \cite{Wysokinski2019} scenarios.
We have found that ferromagnetism in \lcg, due to  different crystal fields on cobalt sites,  predominantly derives from  $d_{xz}$ orbital of Co1 atoms whereas from $d_{xy}$ orbital of Co2 atoms.
In consequence, these two types of atoms develop a difference between net magnetic moments, what  gives  rise to the spatial modulation of a magnetization along $c$-axis; situation, which is formally equivalent to a presence of an antiferromagnetic component. The modulation amplitude,  at ambient pressure rather small, substantially grows with an unit cell contraction, until Co2 atoms becomes nonmagnetic. 
We demonstrate, with a help of a toy-model capturing  found orbital selective ferromagnetic order, that large  amplitude of a spatial modulation is consistent with the recent observation of the resistivity anomaly in   \lcg\ indicating presence of a high-pressure {\it new state} \cite{Canfield2021}. Although, our results indicate that quantum fluctuations do not play a leading role in the interpretation of recent observations in \lcg, we note that the ferromagnetic quantum criticality avoidance can take place at pressure larger than up-to-date measured 5GPa.

\section{Computational details}
 
We have performed DFT calculations by using the VASP package \cite{Kresse93,Kresse96,Kresse96b}. The core and the valence electrons were treated within the Projector Augmented Wave \cite{Kresse99} method with a cutoff of 400 eV for the plane wave basis.
Calculations are done for the system at ambient and applied pressure. Because the experimental atom positions are known only for \lcg\ at ambient pressure \cite{Saunders20} we have performed structural relaxation. For this purpose we have used a revised Perdew-Burke-Ernzerhof functional for solids (PBEsol)  that improves the equilibrium properties of bulk systems\cite{Perdew08,Manca18,Autieri17,Asa18,Asa19}.  Due to the lack of available experimental positions of the atoms in \lcg\ under pressure, we account for applied pressure by the proportional contraction of the total volume, i.e. we keep unchanged ratios between   lattice constants. Choice of PBEsol functional in the studied case of \lcg\ is supported by its good performance   for a structural relaxation at ambient pressure as compared to known atomic positions, a feature that is later demonstrated in Fig. \ref{Magneticmoment} as well as in the Appendix \ref{appA}.

After relaxation, the Local Density Approximation (LDA) with the Perdew–Zunger \cite{Perdew81} parametrization of the Ceperly-Alder \cite{Ceperley80} data have been considered. These calculations have been performed using a  2 $\times$ 12 $\times$ 4 k-point grid, in such a way to have 96 $k$-points in the first Brillouin zone.
For the Brillouin zone integration, we have used the tetrahedron method \cite{Blochl94} in order to get the partial density of states (PDOS) and the magnetic properties of the system. Here, the choice of the different functional for the electronic properties determination is dictated by the DFT community experience with other metallic ferromagnets.
In general, LDA returns smaller magnetic moments as compared to other exchange functionals as shown for instance for elemental Fe, Co or Ni\cite{Fu18,Ek18}.
For metallic magnetic transition-metal compounds without high-spin configuration, PBEsol tends to strongly overestimate the experimental magnetic moment while LDA usually gives a better agreement\cite{Gebhard, Autieri16, Etz12, Campbell21, Autieri17}. 
Recently, the LDA was also used for other Ge-based transition-metal compounds \cite{Grytsiuk21}.
For the sake of completeness, in the  Appendix \ref{appA} we demonstrate a performance of PBEsol for the determination of magnetic moments in \lcg. We show that the mismatch between experimental magnetic moments  and  these obtained by   LDA (cf. Fig. \ref{Magneticmoment}) is  smaller than the one provided by PBEsol.


\section{Structural and electronic properties at ambient pressure}

The crystal structure of La$_5$Co$_2$Ge$_3$ (shown in  Fig. \ref{Structure}) belongs to the R$_5$Co$_2$Ge$_3$ family with the symbol mS40 \cite{Saunders20,Lin17}.
The y coordinates for all atoms in this
structure equal to zero, namely the atoms in this structure
are located either on planes at y = 0 or y = 1/2 arising
from the C center in space group C2/m. The lattice constants are a=18.3540 {\AA}, b=4.3479 {\AA} and c=13.2790 {\AA}.\cite{Saunders20}
Important information for the future discussion is that in \lcg\ there are two crystallographically inequivalent cobalt atoms Co1 and Co2 forming chains along the $b$-axis, as shown in the bottom panel of Fig. \ref{Structure}.

\begin{figure}[t!]
\centering
\includegraphics[width=0.47\textwidth]{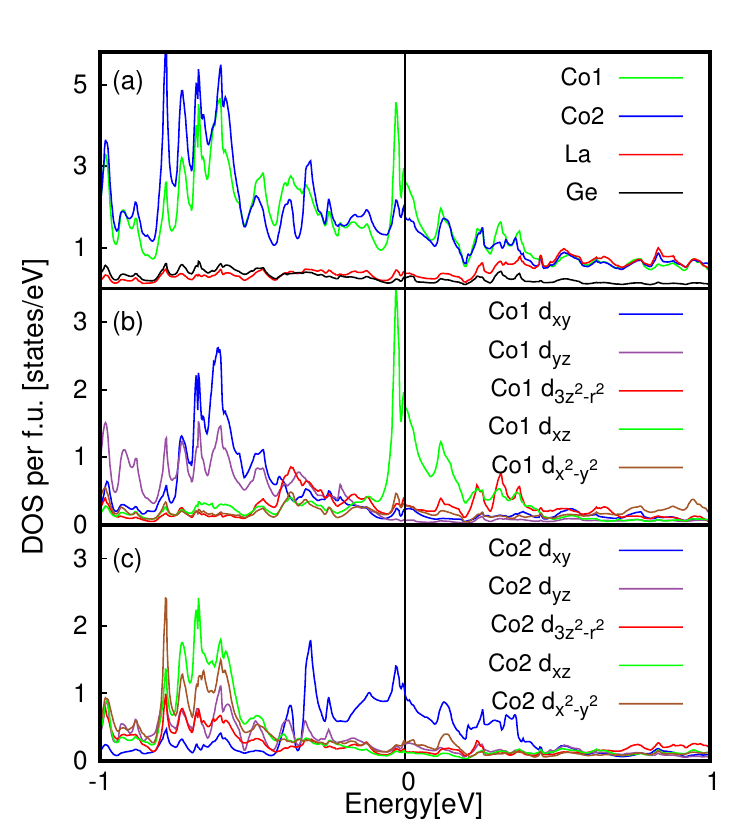}
\caption{(a) Nonmagnetic partial densities of states (PDOS) of $d$-orbitals of  Co1 and Co2 atoms and combined PDOS of $s$, $p$ and $d$ orbitals of La and Ge atoms.
The Fermi level is set at the zero energy.
(b,c) Contributions of the $d$ orbitals of (a) Co1 and (b) Co2 to the nonmagnetic density of states. 
}
\label{DOSNM}
\end{figure}

\begin{figure}[t]
\centering
\includegraphics[width=0.5\textwidth]{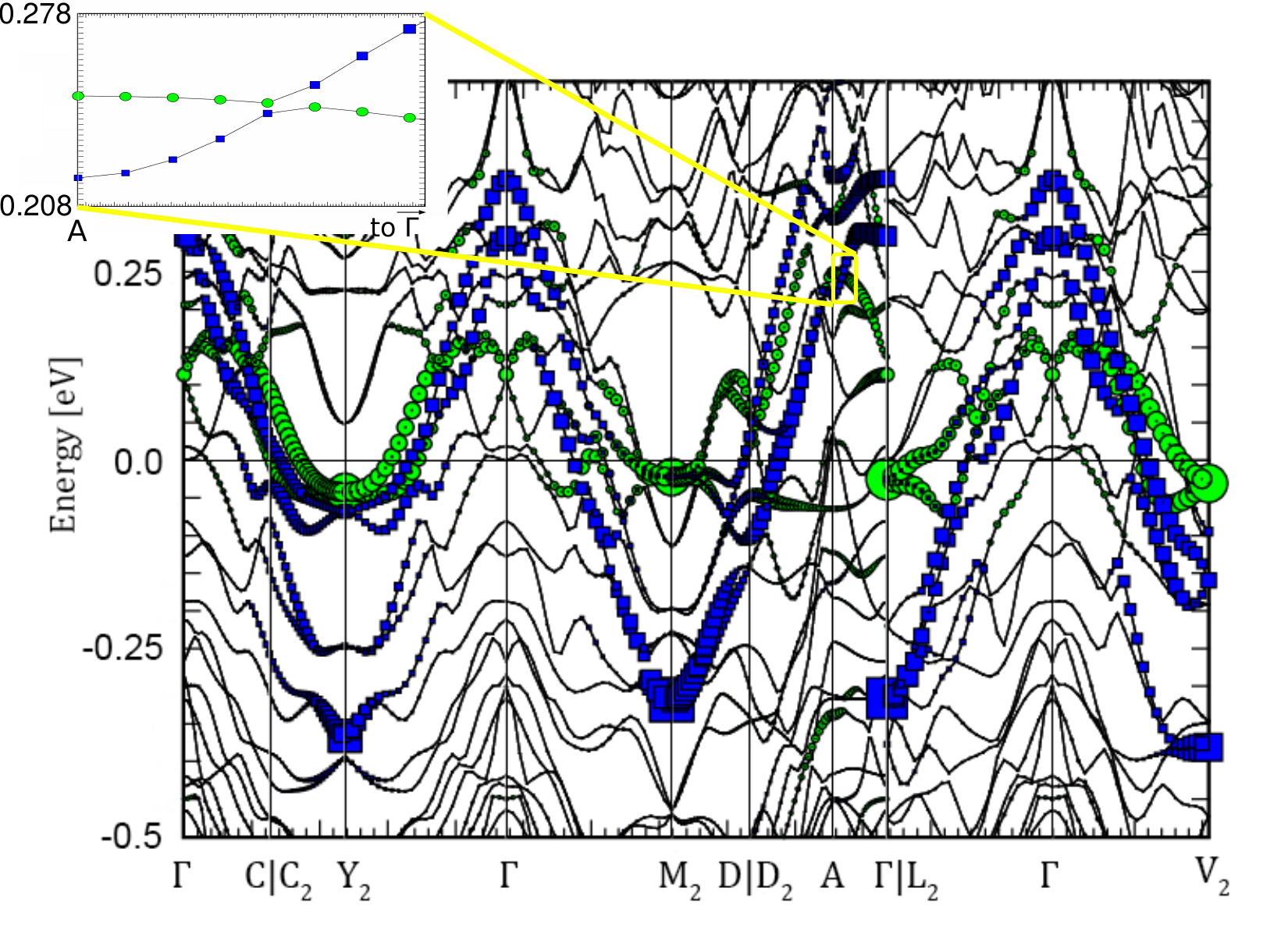}
\caption{Non-magnetic band structure of \lcg\ in the vicinity of the Fermi level with marked contributions from selected $d$ orbitals: d$_{xz}$ of Co1 (green circles) and the d$_{xy}$ of Co2 (blue squares). Size of the symbol correspond to the level of contribution to the particular band. With yellow rectangular we mark exemplary gap (zoomed in the inset) appearing due to hybridization between leading $d$ orbitals along $A$-$\Gamma$ direction (corresponding to $c$-direction in real space).}
\label{NM}
\end{figure}

Regarding electronic properties of \lcg\ we start our analysis by showing in Fig. \ref{DOSNM}a  the partial densities of states  for the dominant orbitals in the vicinity of the Fermi level: $d$ orbitals of Co1 and Co2 and combined $s$, $p$ and $d$ orbitals of La and Ge. It is clear that $d$ orbitals  of Co1 and Co2 atoms by far dominate electronic properties in the vicinity of the Fermi level. Moreover, the calculation shows no charge transfer (no oxidation state), and therefore, due to the crystal field the 4s$^2$3$d^7$ electronic configuration of the elemental Co moves to 3$d^9$. These properties indicate that the low energy states can be reduced to the d-states of the d$^9$ Co atoms, and in result possible magnetic moment in the system  can be equal   1 $\mu_B$/Co (Bohr magneton per Co atom) at maximum.

Cobalt $d$-orbital PDOS in Fig. \ref{DOSNM}a as expected for a metallic Stoner ferromagnet provides a large peak at the Fermi level.
However, different PDOS at the Fermi level for Co1 and Co2  suggest spatially nonuniform magnetic properties of the system, i.e. more fragile magnetism of Co2 than of Co1 atoms. 
In order to have more insight into that feature in Figs. \ref{DOSNM}b,c we plot PDOS contributions from crystal field split $d$ orbitals at Co1 and Co2 atoms. Although all orbitals are mixed, clearly only d$_{xz}$ of Co1 and the d$_{xy}$ of Co2 present a sufficiently large peak at the Fermi level to satisfy Stoner criterion. Therefore, 
in Fig. \ref{NM} we show the non-magnetic band structure in the vicinity of the Fermi level with marked in color contributions from d$_{xz}$ orbital  of Co1 and the d$_{xy}$ orbital  of Co2. Because $d_{xz}$ orbital lies in the plane perpendicular to the $b$-direction, bonding between these orbitals in the Co1 chains is reduced. This is the reason for a  clear difference seen in Fig. \ref{NM} in a width of bands with large contribution from $d_{xz}$ orbital of Co1 and from $d_{xy}$ of Co2 that in consequence leads to distinct peaks at the Fermi level.


Additionally, band structure reflects some of the structural properties of the system. 
The interchain bonding (along $b$-axis) between $d$-orbitals, where distances between neighbouring cobalt sites is the smallest (cf. Fig. \ref{Structure}), is significantly stronger than the intrachain bonding in the $a-c$ plane. 
In this context $A-\Gamma$ path (corresponding to the $c$-direction) is the only one in the whole presented high symmetry path that has no contribution from strongest bonding along the $b$-direction.
Consequently,   Fig. \ref{NM}  shows a sizable dispersion of bands with leading $d$ orbitals contributions along all the paths but $A-\Gamma$.

In the inset of Fig. \ref{NM} we show the exemplary gap formed when    $d_{xz}$ orbital of Co1 and   $d_{xy}$ of Co2 are mixing.  Such gaps, being a measure of hybridization between these orbitals, are of similar order along all paths. However, due to reduced dispersion along $A-\Gamma$ path, bonding between $d_{xz}$ of Co1 and $d_{xy}$ of Co2  should play more important role for the $c$-direction than for the other ones.

\section{Magnetic properties}
 \begin{figure}[t!]
\centering
\includegraphics[width=0.5 \textwidth, trim= 0 0  0 -0.2cm  ]{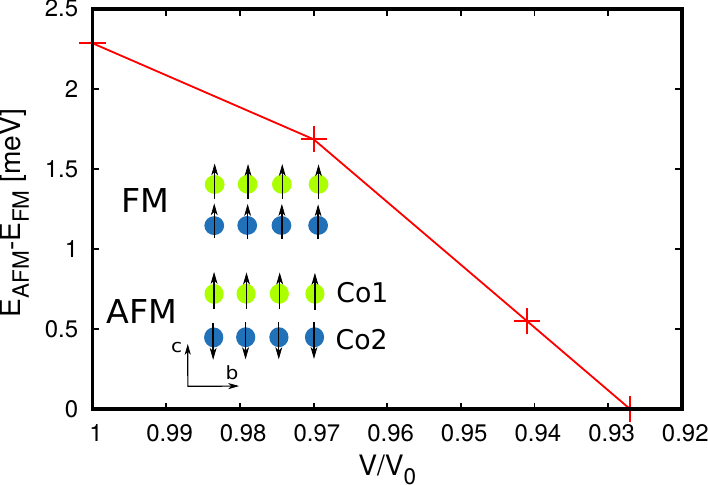}
\caption{Energy difference between realized ferromagnetic state (FM) and the closest in energy antiferromagnetic state (AFM) where moments on Co1 and Co2 atoms are antiparallel (schematically represented in the cartoon on the plot). Zero energy difference takes place because Co2 atoms becomes nonmagnetic and therefore shown in the cartoon AFM order cannot be realized. 
}
\label{energy}
\end{figure}

In this Section, we analyze the magnetic properties of \lcg\ at ambient  and applied pressure. 
Non-magnetic characteristics discussed in the previous Section indicate that magnetic properties of the system are determined predominantly by d$_{xz}$ orbitals of Co1 and the d$_{xy}$ orbitals of Co2 and that they can differ for inequivalent cobalt atoms.  In the following we confirm these expectations.

First, we analyze various magnetic states represented by possible ferromagnetic and antiferromagnetic magnetic moment configurations on cobalt atoms, and find that the system is robustly ferromagnetic. In Fig. \ref{energy} we demonstrate this robustness by showing large  energy difference of the ferromagnetic groundstate to the state with antiferromagnetic alignment  between Co1 chains and Co2 chains at ambient and small applied pressure. In order to instantly avoid confusion, we note that the zero energy difference at applied pressure $V/V_0\simeq 0.93$ is {\bf not} a  phase transition from FM to AFM state but it signals a lost of magnetic moments exclusively at Co2 sites that excludes existence of considered AFM ordering. 
Moreover,  
we find that the magnetic coupling within the chains of Co1 or Co2 atoms along $b$-direction, if present, is by far the strongest one, while these between neighbouring chains, irrespectively whether encompassing  Co1 or Co2 sites, are significantly weaker due to larger atomic distances between cobalt atoms in the $a$-$c$ plane. 

\begin{figure}[t!]
\centering
\includegraphics[width=0.465\textwidth, trim= 0 -0.07cm  0 -0.31cm  ]{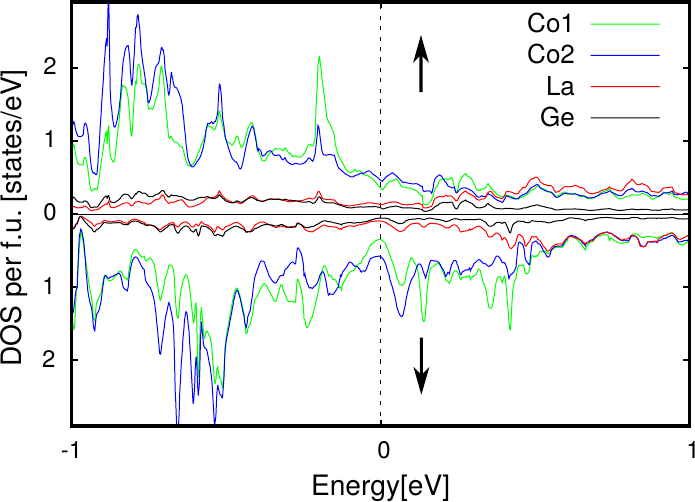}
\caption{Spin resolved PDOS at ambient pressure in ferromagnetic state. PDOS in the vicinity of the Fermi level, similarly as in a non-magnetic state, is dominated by contributions from $d$-orbitals of Co1 and Co2 atoms.
}
\label{DOS_M}
\end{figure}

Additionally, in Fig. \ref{DOS_M} we provide  spin-resolved PDOS of the system at ambient pressure in the ferromagnetic state. It confirms that the material electronic structure in the vicinity of the Fermi level in the magnetic state, similarly as in a non-magnetic state, is dominated by contributions from $d$-orbitals of Co1 and Co2 atoms.

\begin{figure}[t!]
\centering
\includegraphics[trim=0 0 -1cm -0.5cm,clip,width=0.5\textwidth]{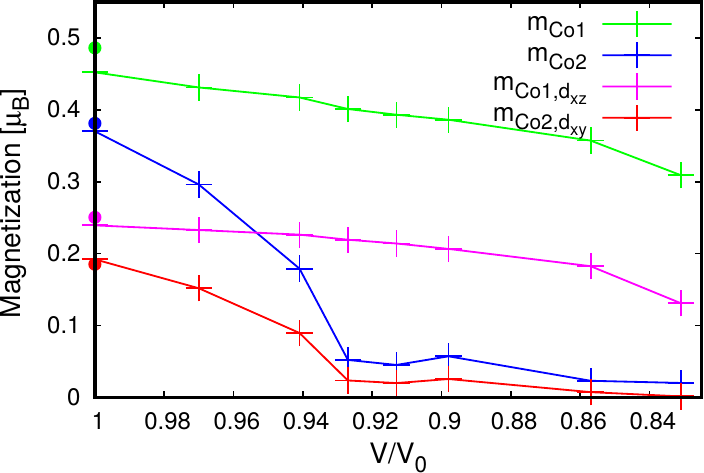}
\caption{Magnetic moments of Co1 and Co2 as a function of decreasing primitive cell volume V/V$_0$, where V$_0$ is the experimental volume at ambient pressure.  The green and blue lines indicate the evolution of magnetic moments at Co1 and Co2 atoms respectively. 
We have also included magnetic moments of the dominant $d$-orbitals,  $d_{xz}$ of Co1 (magenta line) and $d_{xy}$ of Co2 (red line) atoms, providing as much as half of the total magnetization on respective cobalt atoms, confirming their leading role in determining magnetic properties of \lcg.
Magnetic moments marked with cross are obtained with relaxation procedure whereas those marked with a dot are obtained at experimental ambient pressure volume \cite{Saunders20}. 
}
\label{Magneticmoment}
\end{figure}

Next, in Fig. \ref{Magneticmoment} we show  evolution of magnetic moments at Co1 and Co2 sites with applied pressures.  Obtained magnetic moments either for Co1 or Co2 atoms at ambient pressure differ  from experimentally determined net magnetic moment 0.1 $\mu_B$ in \lcg\ \cite{Canfield2021}. However, as these magnetic moments are very low, obtained mismatch is not surprising in a perspective of the known  difficulty of DFT to reproduce the magnetization of itinerant systems not in high-spin configuration (cf. discussion in Section II). Nonetheless, in the following, instead of particular values, we rather focus on the tendencies in which Co1 and Co2 magnetic moments evolve with applied pressure. 

As a rule, applied pressure by simply enlarging electronic bandwidth reduces the magnetic moment in itinerant magnetic systems \cite{Autieri18}. Indeed, behavior of magnetic moments on Co1 and Co2 atoms with decreasing unit cell volume, shown in   Fig. \ref{Magneticmoment} follows this expectation.
However, magnetic moments on Co1 and Co2 sites not only differ already at ambient pressure but they further evolve in a distinct manner with applied pressure. 
Namely, we find that the magnetization on Co2 sites rapidly drops  and Co2 atoms become nonmagnetic for V/V$_0\lesssim0.93$ (residual magnetization is a consequence of a coupling to still magnetic Co1 atoms), while the magnetization of the Co1 at least for V/V$_0\gtrsim0.83$ is rather weakly decreasing. In other words, magnetization difference initially rather small, increases as pressure is applied. In consequence, 
 the system over whole studied pressure range  realizes the same long-range ferromagnetic order varying only with the difference between magnetic moments on Co1 and Co2 sites.

We conclude that \lcg\ is a quasi-one-dimensional orbital-selective ferromagnet. Moreover, because magnetic  properties of the system derive from different $d$-orbitals on Co1 and Co2 sites, magnetic moments on these atoms   evolve independently and in a contrasting manner with applied pressure.
We note, that similarly as for other recently discovered    quasi-one-dimensional compounds \cite{Cuono19,Cuono18,Cuono21a,Cuono21b,Ming18}, 
in \lcg\ weak interchain magnetic couplings are critical for the stabilization of long-range magnetic order in the context of  the Mermin-Wagner theorem \cite{Gelfert01}. 

In the next Section we propose a simple one-dimensional toy model that captures essential features of found orbital selective ferromagnetic order in \lcg\ along $c$-axis, and demonstrate that it is consistent with the recent resistivity measurements \cite{Canfield2021}.


\section{Toy model for spatially modulated ferromagnetism}
\begin{figure}
    \centering
    \includegraphics[trim=0 0 0 0,clip,width=0.48\textwidth]{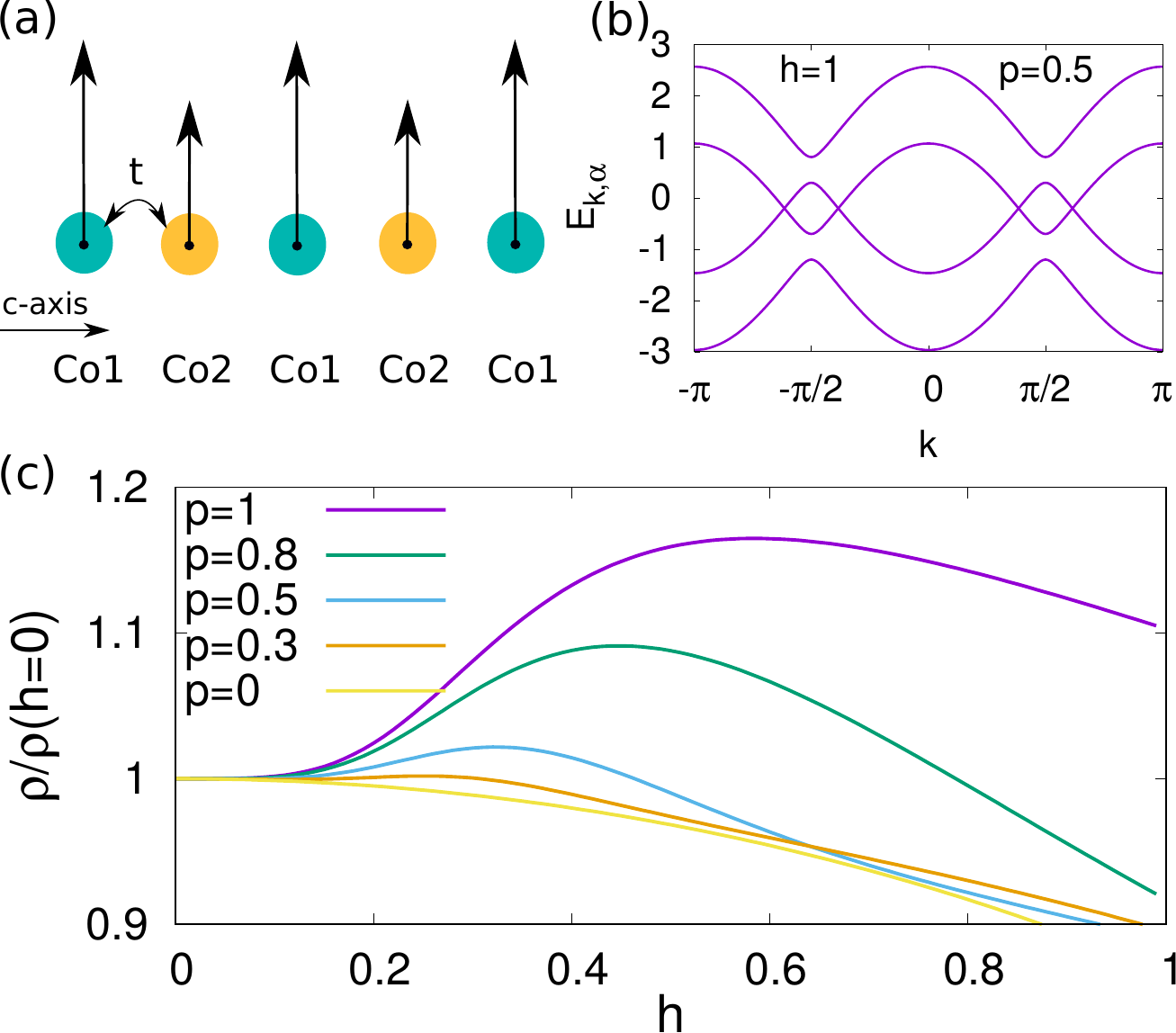}
    \caption{(a) Schematic cartoon of    cobalt sites with different magnetic moments along $c$-axis  visualizing peculiar magnetic behavior of  \lcg\ captured by proposed toy model. (b) Spectrum of the toy model encompassing direct gaps emerging due to nonzero magnetization difference between Co1 and Co2 sites. (c) Normalized resistance as a function of field $h$  for selected values of parameter $p$ corresponding to applied pressure. Chemical potential is set as $\mu=0.2$.  }
    \label{fig_tm}
\end{figure}

We propose a one-dimensional single-orbital toy-model with a site dependent Zeeman field  (cf. Fig. \ref{fig_tm}a for a schematic cartoon),
\begin{equation}
    H=\sum_{k\sigma} (\epsilon_{k}\!-\!\mu) c^\dagger_{k\sigma}c_{k\sigma} -\!\!\!\!\!\sum_{\g i\in {\rm Co1},\sigma}\!\!\!\!\!\sigma h_1\,  c^\dagger_{\g i\sigma}c_{\g i\sigma}-\!\!\!\!\!\sum_{\g i\in {\rm Co2},\sigma}\!\!\!\!\!\sigma h_2\,  c^\dagger_{\g i\sigma}c_{\g i\sigma}
\end{equation}
that mimics spatially modulated ferromagnetism of \lcg\ along $c$-axis. Here, $c_k^{(\dagger)}$ and $c_{\g i}^{(\dagger)}$ are annihilation (creation) operators in the momentum  and real space respectively, $h_1$($h_2$) is a Zeeman field on Co1 (Co2) sites,
$\mu$ is a global chemical potential, $\sigma=\pm$ and a relation dispersion $\epsilon_{k}=-2t\cos k$ is assumed to be generated by a hopping $t$  which is set as an energy unit $t=1$.

Above model, although does not pretend to capture low-energy description of \lcg, does reflect some of the found  features of the compound and the realized magnetic state. For instance, a single-orbital nature of the model is dictated by the fact that  magnetism in \lcg\ is predominantly derived from single $d$ orbital at each cobalt site. 
Moreover, as  dispersion of bands   is sizably reduced along the $c$-axis  in \lcg\ (cf. Fig. \ref{NM}), in the toy model for simplicity metallicity is ensured by a small hopping between neighboring sites  with the same atomic levels, representing crystallographically inequivalent cobalt atoms.
Finally, distinct Zeeman fields, $h_1$ and $h_2$ have a role to generate different magnetic moments on Co1 and Co2 sites respectively according to found by us spatially modulated ferromagnetism  in \lcg.

Additionally, we introduce   parametrization of the Zeeman fields   $h_1=h$ and $h_2=(1-p)h$, $p\in\{0,1\}$. In that manner parameter $p$ controls how fast the difference between magnetic moments increase with an increase of parameter $h$. The increase of   $h$  in turn leads to simultaneous  onset of a magnetization on both type of sites and thus in the first approximation corresponds to a cooling below Curie temperature.
Consequently, $p\simeq0$ corresponds to ambient pressure were for \lcg\ we found small difference between magnetic moments between Co1 and Co2 sites, whereas $p=1$ corresponds to applied pressure where Co2 sites become non-magnetic.  
In the following, we will show that increasing $h$ for $p> 0$  when the difference between magnetic moments becomes sizable,  can lead to the upturn in the resistance  in the  qualitative agreement with observations for \lcg\ in the {\it new state} \cite{Canfield2021}.

Fourier transform of site-dependent Zeeman splitting leads to Hamiltonian in reduced Brillouin zone (RBZ) that formally describes the coexistence of ferromagnetic and antiferromagnetic orders,
\begin{equation}
    H=\!\!\!\!\sum_{  k\in {\rm RBZ},\sigma}\!\!\!\!\!\!\Psi_{k\sigma}^\dagger\!\!\begin{pmatrix}
    \epsilon_k\!-\!\mu\!-\! \sigma h_{FM}& -\sigma h_{AFM}\\
    - \sigma h_{AFM} & \!- \epsilon_k\!-\!\mu\!-\! \sigma h_{FM}
    \end{pmatrix}\!\!\Psi_{k\sigma},
\end{equation}
where $h_{FM}=h(1-p/2)$, $h_{AFM}=h\, p/2$  and $\Psi_{k\sigma}^\dagger\equiv\{c^\dagger_{k \sigma},c^\dagger_{k+\pi \sigma}\}$.
The spectrum of the Hamiltonian consists of four eigenvalues $E_{k,\{1-4\}}=\pm 
h_{FM}-\mu\pm\sqrt{h_{AFM}^2+\epsilon_k^2}$. Resulting spectrum for $p>0$ encompasses direct gaps in the same spin channels due to the spatial modulation of magnetic moments (cf. Fig. \ref{fig_tm}b). 
In order to calculate coherent conductance at zero temperature with respect to increasing $h$ with the  toy model's spectrum we use Landauer formula,
\begin{equation}
   G=\frac{dI}{dV}= \frac{e^2}{h}T(0)
\end{equation}
 where transmission function $T$  is related to the spectral function of the model and can be approximated as
 \begin{equation}
     T(E)=\sum_{k\alpha}\frac{1}{\pi}\frac{\Gamma}{(E-E_{k\alpha})^2+\Gamma^2}\,\,\,\,\, .
 \end{equation}
 Here factor $\Gamma$ broadens spectral function in order to account for a realistic nature of the system encompassing various sources of disorder and scattering processes, and we assume $\Gamma=0.2$.  
 
 In Figure \ref{fig_tm}c we plot   resistance, $\rho=1/G$  as a function of field $h$ and  normalized to $\rho(h\!\!=\!\!0)$ for selected values of  parameter $p$ and for global chemical potential set to $\mu=0.2$.
 At sizable values of parameter $p>0.3$ resistance shows a clear upturn  with increasing $h$, in the  qualitative agreement with observations for \lcg\ in the {\it new state} \cite{Canfield2021}.
Here, we would like to underline two features resulting from the above analysis:   the magnetization difference between Co1 and Co2 sites needs to be larger than some critical value to produce upturn in the resistivity,  and that   Co2 sites do not need to be  necessarily nonmagnetic ($p=1$) to lead to a visible increase in resistivity. 

\section{Discussion}
Our DFT results supported with the analysis of the toy model suggest that
the increasing magnetic moment difference between Co1 and Co2 atoms can be responsible for the recently observed anomalous behavior of the resistivity along $c$-axis in \lcg\ \cite{Canfield2021}. In the following, we put all our findings together into a coherent interpretation.

The magnetization difference between Co1 and Co2 sites, present even at ambient pressure, follows that magnetic state in \lcg\ 
is spatially modulated {\bf only} along the $c$-axis due to the structural alternation of these atoms  (cf. Fig. \ref{Structure}).
Difference between magnetic moments between Co1 and Co2 atoms, at ambient pressure rather small, drastically increases with the unit cell contraction until Co2 atoms become nonmagnetic at V/V$_0\simeq0.93$ (cf. Fig. \ref{Magneticmoment}), when it starts to weakly decrease.

Such spatial modulation, although it has structural origin, is formally equivalent to the presence of  an antiferromagnetic component on top of uniform ferromagnetic state. Therefore, as we demonstrated with the help of the toy model, it is able to produce a  direct gaps in the spectrum (cf. Fig. \ref{fig_tm}b). These gaps are small at ambient pressure and thus likely hindered by various sources of disorder and scattering processes in the transport measurements. On the other hand, at applied pressure, once  the 
 difference between magnetic moments on Co1 and Co2 sites becomes sizable, direct gaps due to spatial modulation of magnetization can give rise to the  increase of the resistivity when cooling below Curie temperature (cf. Fig. \ref{fig_tm}c - upturn is visible for $p\gtrsim0.3$) what resembles resistivity anomaly in \lcg\ \cite{Canfield2021}. 
Because the  difference between magnetic moments at Co1 and Co2 sites rapidly increases with applied pressure (cf. Fig. \ref{Magneticmoment})  we suggest that the {\it new phase} in \lcg\ is  characterized by nonmagnetic Co2 atoms when  spatial modulation of the magnetization is simply the largest.
In summary, our interpretation based on orbital selective ferromagnetism suggests that the {\it new state} in fact represents  the same ferromagnetic order realized by \lcg\ at ambient pressure, and the appearance of the resistivity anomaly signals a crossover instead of a phase transition.

For the sake of completeness in Appendix \ref{appB} we show 
  band structure plots in a magnetic state at ambient and applied pressure with marked in color contributions from the dominant $d$-orbitals,  $d_{xz}$ of Co1  and $d_{xy}$ of Co2. In band structure plots we are able to identify gaps formed at the crossing of bands with major contribution from these orbitals, and they are of similar size as for the non-magnetic state (cf. Fig. \ref{NM}). 
  However, given the gap sizes, and the fact that gaps due to spatial modulation of moments and due to hybridization take place in the same spin channel,  it is impossible  to unambiguously identify contributions of magnetic origin.

Our interpretation of spatial modulation of magnetization  in \lcg\ is related to structural features of the material, in contrast to purely electronic in origin, quantum criticality avoidance \cite{Chubukov2004,Conduit2009,Karahasanovic2012} or correlated-uncorrelated orbital mixing \cite{Wysokinski2019} scenarios. However, although  electronic properties  of \lcg\   definitely excludes correlated-uncorrelated orbital mixing \cite{Wysokinski2019} approach, the effects related to quantum fluctuation \cite{Belitz1997} can be still of critical importance at larger than up to date measured (5GPa) pressure where ferromagnetism of Co1 would go down towards zero temperature.

 \section{Summary}
 
 In this work, we study electronic and magnetic properties of metallic low-$T_C$ ferromagnet \lcg\ at ambient and applied pressure   by means of density functional calculations. We establish that \lcg\ is quasi-one-dimensional orbital selective ferromagnet. Namely, we found that due to crystal field effects two different $d$ orbitals  on two crystallographically inequivalent cobalt sites Co1 and Co2 are predominantly responsible for emergence of ferromagnetism. In consequence, magnetic moment on each type of  cobalt atoms in response to applied pressure  evolves  in a drastically different manner: Co2 atoms for relatively small unit cell contraction become nonmagnetic in sharp contrast to rather robust ferromagnetism of Co1 atoms. The alternating sequence of Co1 and Co2 atoms   provides spatial modulation of magnetization, situation, though having structural origin, is  formally equivalent to the presence of an antiferromagnetic component on top of ferromagnetic state. 
 
 Relying on these findings and analysis of a simple toy model mimicking found magnetic order, we argue that recently observed    {\it new state}  in \lcg\ \cite{Canfield2021} is related to the  sizable increase with applied pressure of the amplitude of a spatial modulation of  magnetic moments on cobalt sites; situation that can give rise to the upturn in the resistivity when cooling below Curie temperature. 
 Our results indicate that  the {\it new state}  in \lcg\ is the same ferromagnetic order as this realized at ambient pressure and the appearance of a resistivity anomaly signals a crossover instead of a phase transition. 
 
 Finally, we note that our calculations indicate orbital selectivity rather than quantum critical avoidance as a driving mechanism for the presence of a spatial modulation of magnetic moments in \lcg. Nevertheless,  effects related to quantum fluctuation \cite{Belitz1997} can be still of critical importance at larger than up to date measured (5GPa) pressures where ferromagnetism of Co1 would go down towards zero temperature.

 \section*{Acknowledgments}
 We are grateful to Paul Canfield and members of his team for discussion on the subtleties and consequences of the experimental measurements on \lcg.
The work is supported by the Foundation for Polish Science through the International Research
Agendas program co-financed by the European Union within the Smart Growth Operational Programme.
We acknowledge the access to the computing facilities of the
Interdisciplinary Center of Modeling at the University of
Warsaw, Grants No. G75-10, G84-0, GB84-1 and GB84-7.
We acknowledge the CINECA award under the
ISCRA initiatives IsC81 "DISTANCE" and  IsC85 "TOPMOST"
Grant, for the availability of high-performance computing
resources and support.

\appendix
\section{PBEsol vs LDA functional}
\label{appA}
\begin{figure}[t!]
\centering
\includegraphics[width=0.49\textwidth]{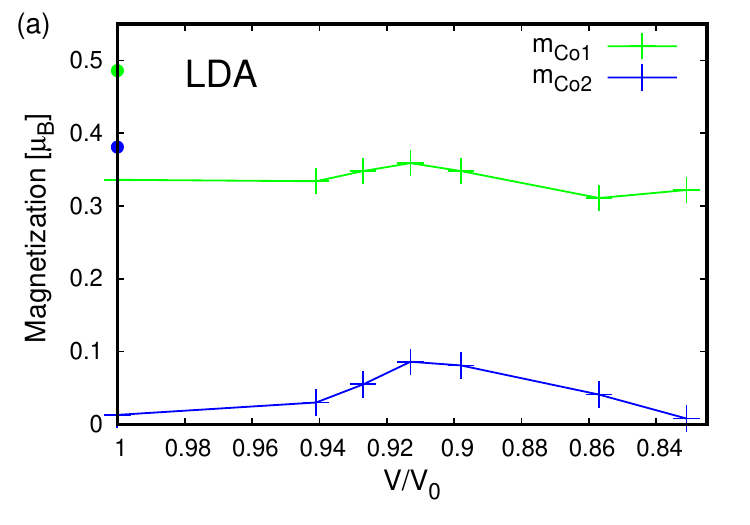}
\includegraphics[width=0.49\textwidth]{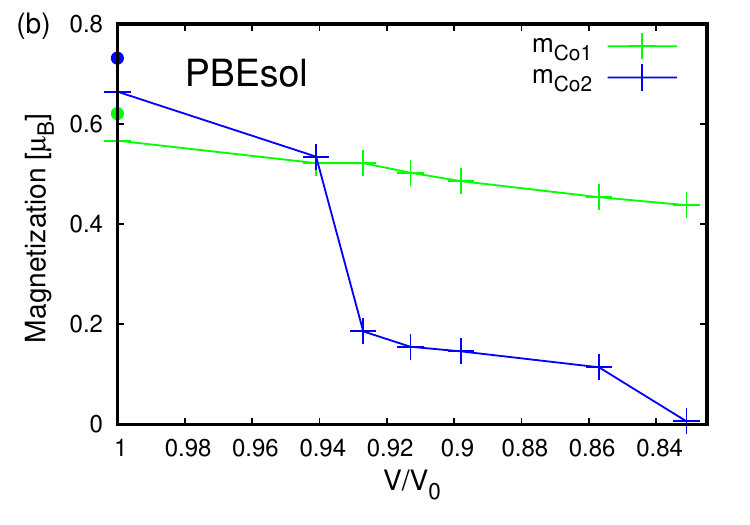}
\caption{Magnetic moments of Co1 and Co2 as a function of decreasing primitive cell volume V/V$_0$, where V$_0$ is the experimental volume at ambient pressure given that structural relaxation is performed and electronic properties determined with the same functional:  (a) LDA and (b) PBEsol.  
Magnetic moments marked with cross are obtained with relaxation procedure whereas those marked with a dot are obtained at experimental ambient pressure volume \cite{Saunders20}. 
}
\label{Magneticmoment_app}
\end{figure}    

\begin{figure*} 
\centering
\includegraphics[width=\columnwidth,   angle=0]{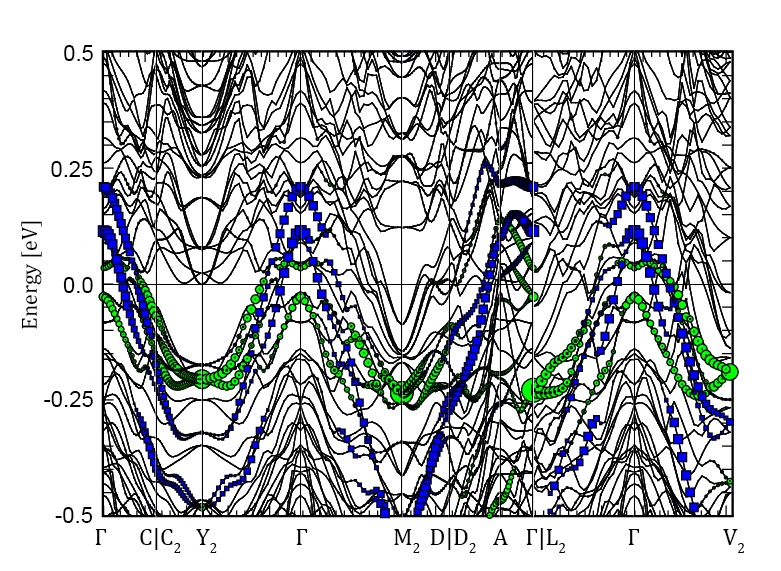}
\includegraphics[width=\columnwidth,   angle=0]{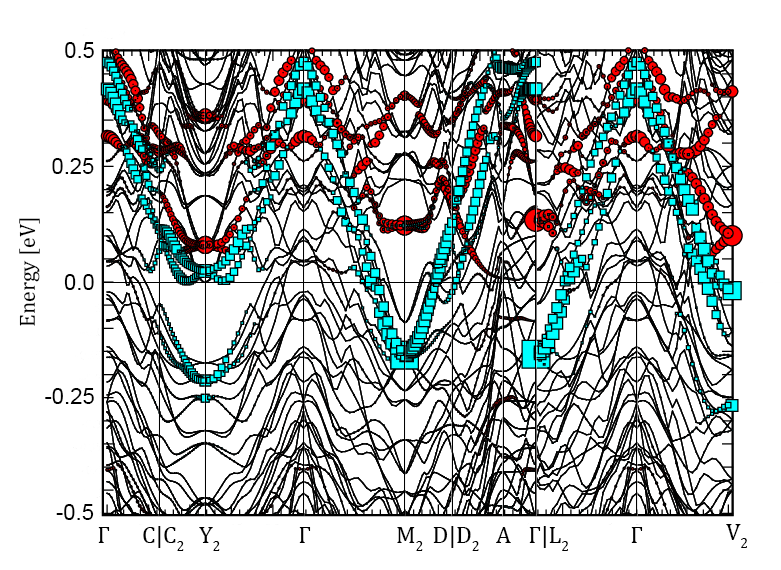}
\caption{Magnetic band structure of \lcg\ at ambient pressure in the vicinity of the Fermi level.
The left panel shows bands in the spin up channel  (green circles indicates d$_{xz}$ orbital of Co1 contribution,  blue squares: d$_{xy}$ of Co2), and right panel shows bands in the spin down channel  (red circles: d$_{xz}$ of Co1, cyan squares: d$_{xy}$ of Co2).  Size of the color symbol corresponds to the level of contribution to the particular band.}
\label{Magnetic_100}
\end{figure*}

\begin{figure*}
\centering
\includegraphics[width=\columnwidth,   angle=0]{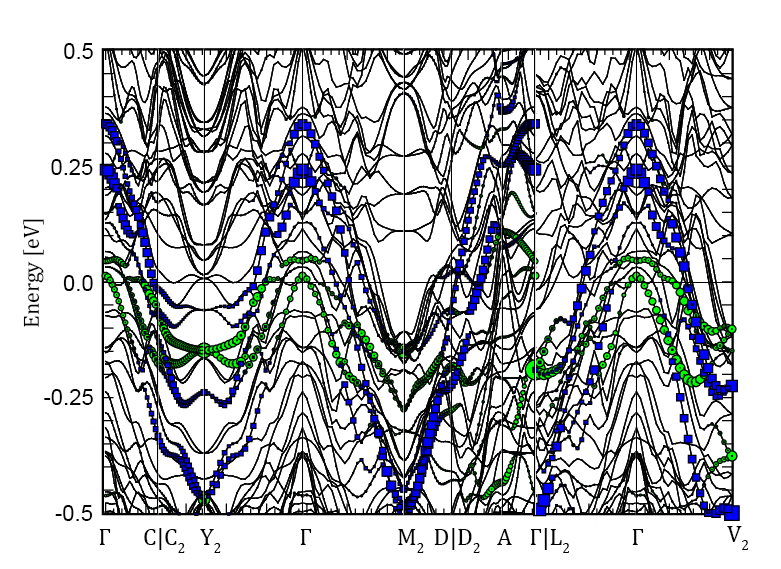}
\includegraphics[width=\columnwidth,   angle=0]{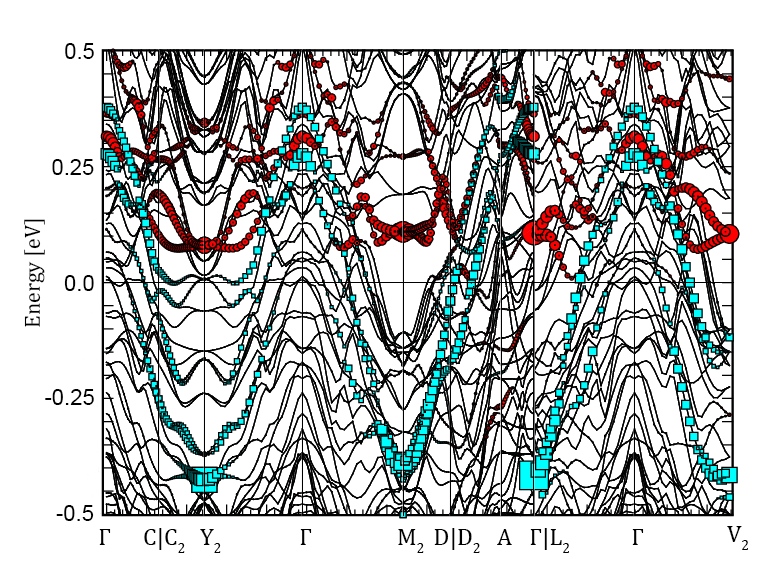}
\caption{Magnetic band structure of \lcg\ at applied pressure, V/V$_0$=0.913 in the vicinity of the Fermi level.
The left panel shows bands in the spin up channel  (green circles indicates d$_{xz}$ orbital of Co1 contribution,  blue squares: d$_{xy}$ of Co2), and right panel shows bands in the spin down channel  (red circles: d$_{xz}$ of Co1, cyan squares: d$_{xy}$ of Co2).  Size of the color symbol corresponds to the level of contribution to the particular band.}
\label{Magnetic_0913}
\end{figure*}

Here we support our choice of PBEsol functional for relaxation and LDA functional for determination of electronic properties.  First, we have checked consistent performance of the structural relaxation within LDA. In Fig. \ref{Magneticmoment_app}a we show obtained magnetic moments for Co1 and Co2 atoms. Poor performance of LDA for the determination of structural relaxation is clear by the huge mismatch between magnetic moments at ambient pressure obtained with relaxed atomic positions and experimental ones. On the contrary, calculations with PBEsol functional used for relaxation, as it is demonstrated by Fig. \ref{Magneticmoment_app}b (as well as by Fig. \ref{Magneticmoment}  in the main text) provides very good agreement between magnetic moments at ambient pressure obtained with relaxed atomic positions and experimental ones.

Having established that PBEsol functional is better suited for a structural relaxation it remains to determine whether PBEsol is better or worse than LDA functional in evaluation of electronic properties. Here, we  use the criterion which of the two provide  smaller mismatch between obtained and measured magnetic moments. Experimentally measured magnetic moment per cobalt atom is around 0.1$\mu_B$. Although both functionals overestimate magnetic moments at ambient pressure,  LDA (cf. Fig. \ref{Magneticmoment}) provides smaller mismatch  than PBEsol (cf. Fig. \ref{Magneticmoment_app}b). Nevertheless, we note that qualitative behavior of magnetization on inequivalent cobalt sites with applied pressure is very similar for both functionals. Namely that Co1 and Co2 sites develop distinct magnetic moments and that magnetization difference between these sites substantially increases at   applied pressure.    Therefore we underline that provided by us interpretation of the {\it new state} in \lcg\ is consistent independently to the choice of  functional for electronic properties determination.

\section{Band structure in magnetic state}\label{appB}

In this appendix, for the sake of completeness, we report the band structure in the magnetic state. In Fig. \ref{Magnetic_100} we show spin-resolved  bands (left panel - spin up, right panel - spin down) in the magnetic state at ambient pressure, where   with different colors we highlight  d$_{xz}$ of Co1 and d$_{xy}$ of Co2 orbital contributions.
The same type of band structure plot is shown in Fig. \ref{Magnetic_0913} with a difference that it represents system under pressure,  V/V$_0$=0.913, where Co2 sites are non-magnetic.


 \end{document}